\documentclass{article}

    \usepackage[nonatbib,final]{neurips_2021_ml4ps}

\usepackage[square,numbers]{natbib}
\usepackage[utf8]{inputenc} 
\usepackage[T1]{fontenc}    
\usepackage{hyperref}       
\usepackage{url}            
\usepackage{booktabs}       
\usepackage{amsfonts}       
\usepackage{mathtools}
\usepackage{nicefrac}       
\usepackage{microtype}      
\usepackage[pdftex]{graphicx}
\usepackage{bm}
\usepackage{braket}
\usepackage{multirow}

\usepackage{color}
\usepackage{xcolor}         

\title{Supplementing Recurrent Neural Network Wave Functions with Symmetry and Annealing to Improve Accuracy}

\author{%
  Mohamed Hibat-Allah$^{1,2}$, Roger G. Melko$^{2,3}$, Juan Carrasquilla$^{1,2}$ \\
  $^1$ Vector Institute, MaRS  Centre,  Toronto,  Ontario,  M5G  1M1,  Canada\\
   $^2$ Department of Physics and Astronomy, University of Waterloo, Ontario, N2L 3G1, Canada \\
  $^3$ Perimeter Institute for Theoretical Physics, Waterloo, ON N2L 2Y5, Canada \\
   \texttt{mohamed.hibat.allah@uwaterloo.ca} 
}

\begin{document}

\maketitle

\begin{abstract}
Recurrent neural networks (RNNs) are a class of neural networks that have emerged from the paradigm of artificial intelligence and has enabled lots of interesting advances in the field of natural language processing. Interestingly, these architectures were shown to be powerful ansatze to approximate the ground state of quantum systems~\cite{RNNWF}. Here, we build over the results of Ref.~\cite{RNNWF} and construct a more powerful RNN wave function ansatz in two dimensions. We use symmetry and annealing to obtain accurate estimates of ground state energies of the two-dimensional (2D) Heisenberg model, on the square lattice and on the triangular lattice. We show that our method is superior to Density Matrix Renormalisation Group (DMRG) for system sizes larger than or equal to $14 \times 14$ on the triangular lattice.
\end{abstract}

\section{Introduction}

In recent years, deep learning has proven useful in quantum many-body physics~\cite{androsiuk1993,LAGARIS19971,carrasquilla2017nature,Carleo_2017}. Namely in tasks such as quantum state tomography~\cite{torlainqs2018,Carrasquilla_2019} and finding ground states of quantum many-body Hamiltonians~\cite{zi2018, Di_Luo, Pfau_2020, Hermann_2020, choo_fermionicnqs2020, Sharir_2020, RNNWF, roth2020iterative}. Yet, up to now, the numerical study of frustrated quantum systems remains challenging due to the debilitating sign problem~\cite{Troyer_2005,westerhoutGeneralizationPropertiesNeural2020}.

Interestingly, recurrent neural networks (RNNs)~\cite{hochreiter1997long,graves2012supervised,lipton2015} have shown a significant advantage in the study of quantum systems~\cite{Carrasquilla_2019,RNNWF, roth2020iterative, casert2020dynamical}. In this paper, we aim to show the potential of RNNs in the study of frustrated quantum systems that suffer from the sign problem~\cite{Troyer_2005,westerhoutGeneralizationPropertiesNeural2020}. To do this,  we focus on the prototypical Heisenberg model and show that by applying symmetries and annealing leads to accurate representations of the ground state. In Sec.~2, we describe our improved RNN architecture and annealing scheme. In Sec.~3, we show results for the Heisenberg model on the square and the triangular lattice. We demonstrate that our method is superior compared to the Density Matrix Renomalization Group (DMRG)~\citep{White1992, Schollw_ck_2005} on system sizes larger than $14 \times 14$ up to $16 \times 16$. 


\section{Methods}
\subsection{Recurrent neural network wave functions}
\label{sec:RNNWF}

To model a complex-valued wave function $\ket{\Psi}$, we use the following decomposition
\begin{equation*}
   \Psi_{\bm{\theta}}(\bm{\sigma}) = \sqrt{p_{\bm{\theta}}(\bm{\sigma})} \exp(\text{i} \phi_{\bm{\theta}}(\bm{\sigma})),
\end{equation*}
where $\bm{\theta}$ denotes the variational parameters of the ansatz wave function $\ket{\Psi_{\bm{\theta}}}$, and $\bm{\sigma} = (\sigma_1, \sigma_2, \ldots, \sigma_N)$ is a spin configuration. In our paper, we use a complex RNN (cRNN) wave function to handle states that change sign in the computational basis~\cite{RNNWF}. A key feature of RNN wave functions is the ability to obtain uncorrelated samples to estimate observables through autoregressive sampling~\cite{RNNWF, Bengio2000}. More details about this sampling procedure in 1D and 2D can be found in Refs.~\cite{RNNWF, VNA}.

\begin{figure}
    \centering
    \includegraphics[width = \linewidth]{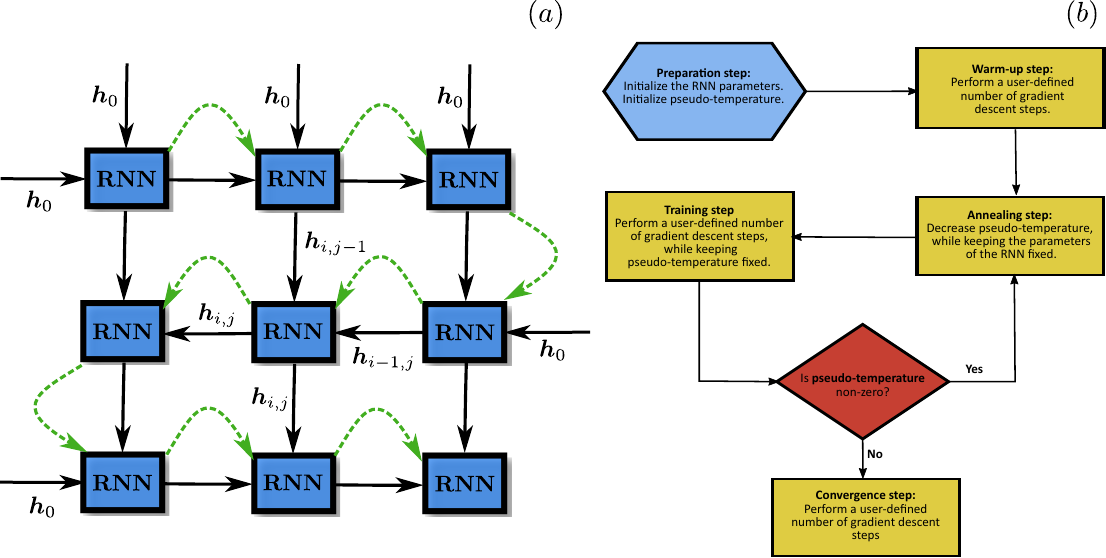}
    \caption{(a) A graphical illustration of a 2D RNN. Each RNN cell receives two hidden states $\bm{h}_{i,j-1}$ and $\bm{h}_{i-1,j}$, as well as two input vectors $\bm{\sigma}_{i,j-1}$ and $\bm{\sigma}_{i-1,j}$ (not shown) as illustrated by the black arrows. The green arrows correspond to the zigzag path we use for 2D autoregressive sampling. The initial memory state $\bm{h}_0$ of the RNN and the initial inputs $\bm{\sigma}_0$ (not shown) are null vectors. (b) A flowchart describing our annealing implementation.}
    \label{fig:methods}
\end{figure}

In order to efficiently encode 2D correlations, we use a two-dimensional recursion relation as follows~\cite{VNA}:
\begin{equation}
    \bm{h}_{i,j} = F\!\left(
    [\bm{\sigma}_{i-1,j} ; \bm{\sigma}_{i,j-1}]T 
    [\bm{h}_{i-1,j} ; \bm{h}_{i,j-1}] + \bm{b}\right).
    \label{eq:2DTRNN}
\end{equation}
$\bm{h}_{i,j}$ is a hidden state with two indices for each spin in the 2D lattice. The symbol $[. ; .]$ denotes concatenation operation of two vectors, while $\bm{\sigma}_{i,j}$ is the one-hot encoding of the spin $\sigma_{i,j}$. The tuneable arrays $T$ and $\bm{b}$ are a tensor and bias vector, respectively. Since $\bm{h}_{i,j}$ can be considered a summary of the history of spins that were previously generated, it is used to compute the conditional probability and phase as follows:
\begin{align}
    p_{\bm{\theta}}(\sigma_{i,j}| \sigma_{<i,j}) &= \text{Softmax}(U \bm{h}_{i,j} + \bm{c}) \cdot \bm{\sigma}_{i,j}, 
    \label{eq:prob_softmax}
    \\
    \phi_{\bm{\theta}}(\sigma_{i,j}| \sigma_{<i,j}) &= \pi \ \text{Softsign}(V \bm{h}_{i,j} + \bm{d}) \cdot \bm{\sigma}_{i,j}.
    \label{eq:phase_softsign}
\end{align}
Here $\sigma_{<i,j}$ stands for the spins that were previously generated on the zigzag path as shown in Fig.~\ref{fig:methods}(a). $U,V$ and $\bm{c}, \bm{d}$ are additional variational parameters. From the two previous relations, the probability $p_{\bm{\theta}}(\bm{\sigma})$ and the phase $\phi_{\bm{\theta}}(\bm{\sigma})$ can be computed respectively by taking the product (resp. sum) of $p_{\bm{\theta}}(\sigma_{i,j}| \sigma_{<i,j})$ (resp. $\phi_{\bm{\theta}}(\sigma_{i,j}| \sigma_{<i,j})$). 

We note that to compute $\bm{h}_{i,j}$ in the recursion relation~\eqref{eq:2DTRNN}, we use the neighbouring hidden states and spins, as demonstrated in Fig.~\ref{fig:methods}(a). We also remark that a tensorized operation of the hidden states and the spins is chosen instead of a concatenation operation (see Eq.~\eqref{eq:2DTRNN}) to improve the expressivity of the 2D RNN as shown in Ref.~\cite{VNA}. In this paper, we also use a gated version of the 2D RNN to further improve the accuracy of our 2D cRNN wave function ansatz. More details are shown in Appendix.~\ref{sec:GRU}.

Furthermore, symmetries in a system of interest can be incorporated in RNN wave functions, namely $U(1)$ symmetry and discrete lattice symmetries as it was shown in Ref.~\cite{RNNWF}. We highlight that adding symmetries significantly improves the accuracy of a variational calculation, as motivated in the literature with the example of RBM wave functions~\cite{Nomura_2021}. We finally emphasize that the variational parameters $\bm{\theta}$ are optimized with the variational Monte Carlo (VMC) scheme to minimize $E_{\bm{\theta}} = \bra{\Psi_{\bm{\theta}}} \hat{H} \ket{\Psi_{\bm{\theta}}}$ where $\hat{H}$ is the Hamiltonian describing a quantum system of interest. More details about this scheme can be found in the Appendix of Ref.~\cite{RNNWF}.

\subsection{Variational Annealing}
\label{sec:annealing}

Finding the ground state of a non-stoquastic Hamiltonian is limited by the sign problem~\cite{Troyer_2005,westerhoutGeneralizationPropertiesNeural2020}, and has shown to induce rugged landscapes during the VMC optimization~\cite{Bukov_2021}. The latter can make our ansatz stuck in a local minima, similarly to what is commonly observed in spin glass models at low temperatures. We take inspiration from the concept of annealing, where we supplement VMC with an artificial temperature to overcome local minima during a VMC optimization. This procedure has been already explored in Ref.~\cite{VNA} for the purpose of finding ground states of spin-glass models. Here, we explore this idea in the context of finding ground states of frustrated quantum systems.

In order to supplement the VMC scheme with annealing, we add a pseudo-entropy~\cite{roth2020iterative} on top of variational energy $E_{\bm{\theta}}$ to obtain a variational free energy:
\begin{equation}
    F_{\bm{\theta}}(n) = E_{\bm{\theta}} - T(n) S_{\rm classical} ( p_{\bm{\theta}} ),
    \label{eq:FreeEnergy}
\end{equation}
where $S_{\rm classical}$ is the von Neumann entropy, given by
\begin{equation}
    S_{\rm classical} (p_{\bm{\theta}}) = - \sum_{\bm{\sigma}} p_{\bm{\theta}}(\bm{\sigma}) \log\left(p_{\bm{\theta}}(\bm{\sigma})\right).
    \label{eq:vnEntropy}
\end{equation}
Here the sum goes over all possible configurations $\{\bm{\sigma}\}$. In this case $T(n)$ is a pseudo-temperature. The latter is annealed starting from some initial temperature $T_0$ to zero temperature as follows: $T(n) = T_0 (1-n/N_{\rm annealing})$ where $n \in [ 0, N_{\rm annealing} ]$ and $N_{\rm annealing}$ is the total number of annealing steps. 

In our setup, we follow the same scheme in Ref.~\cite{VNA}, where we first do $N_{\rm warmup}$ training steps to prepare our RNN wave function close to the minimum of $F_{\bm{\theta}}(n = 0)$. In the second step, we keep the parameters of our ansatz fixed and we change $n$ to $n+1$. In step 3, we perform $N_{\rm train}$ gradient steps to bring the RNN ansatz closer to the minimum of $F_{\bm{\theta}}(n+1)$. After iterating over steps 2 and 3, we arrive at $n = N_{\rm annealing}$. At this point, we perform a user-defined number of gradient steps $N_{\rm convergence}$ to further improve our estimate of the ground state energy of $\hat{H}$. A flowchart detailing this procedure is provided in Fig.~\ref{fig:methods}(b).

\section{Results}
\label{sec:results}

We focus our attention on the task of finding the ground state of the two dimensional anti-ferromagnetic Heisenberg model on the square lattice and on the triangular lattice with open boundary conditions (OBC). The Hamiltonian is given as follows:
\begin{equation*}
    \hat{H} = \frac{1}{4} \sum_{\langle i,j \rangle} ( \hat{\sigma}_i^{x} \hat{\sigma}_j^{x} + \hat{\sigma}_i^{y} \hat{\sigma}_j^{y} + \hat{\sigma}_i^{z} \hat{\sigma}_j^{z}),
\end{equation*}
where the sum is over nearest neighbours and $\hat{\sigma}_i^{x,y,z}$ are Pauli matrices. In the square lattice case, $\hat{H}$ has a $C_{4v}$ symmetry\footnote{A point group with four rotations and four mirror reflections (vertical, horizontal and diagonal).}, whereas for the case of the triangular lattice, $\hat{H}$ has a $C_{2d}$ symmetry\footnote{A point group with two rotations and two diagonal mirror reflections.}. We also remark that in both cases, the ground state has zero magnetization~\cite{Lieb1962}, due to the $U(1)$ symmetry of the Hamiltonian $\hat{H}$. We show that enforcing these symmetries in our ansatz allows to obtain a better accuracy without adding more parameters. More details about how to apply symmetries to the 2D cRNN ansatz can be found in the Appendix of Ref.~\cite{RNNWF}. 

\subsection{Heisenberg model on the square lattice}

Since the Hamiltonian $\hat{H}$ can be made stoquastic on the sqaure lattice~\cite{Marshall1955, CAPRIOTTI_2001}, we set $T_0 = 0$, i.e., we do not use annealing in this case. The latter choice is motivated by the observations shown in Appendix.~\ref{sec:annealing}. 

\begin{figure}
    \centering
    \includegraphics[width = 0.95\linewidth]{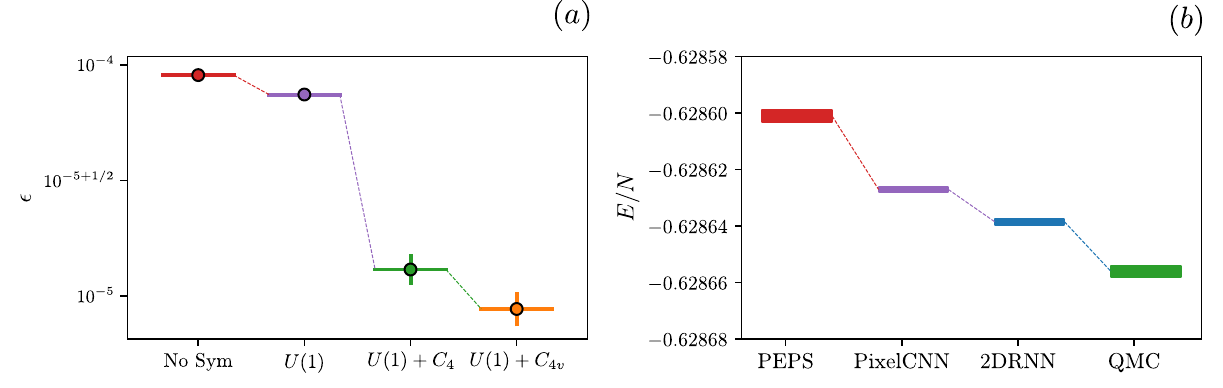}
    \caption{(a) A plot of the relative error $\epsilon$ after applying different symmetries of the Heisenberg model on the square lattice with size $6 \times 6$. The relative error is computed with respect to the DMRG energy~\cite{roth2020iterative}. $C_4$ is the point group of four rotations. (b) A comparison of the energy per site obtained with our 2DRNN ansatz and PEPS~\cite{Liu_2017}, PixelCNN~\cite{Sharir_2020} and QMC~\cite{Liu_2017} on the Heisenberg model on the square lattice with size $10 \times 10$.}
    \label{fig:2DHeis_square}
\end{figure}

First, we train our 2D cRNN wave function to find the ground state of the $6 \times 6$ square lattice with and without applying the symmetries of the Hamiltonian $\hat{H}$. We find in Fig.~\ref{fig:2DHeis_square}(a) that increasing the number of symmetries encoded in the cRNN leads to a more accurate estimation of the ground state energy.

We also use our 2D cRNN wave function to estimate the ground state of the $10 \times 10$ square lattice after applying $U(1)$ and $C_{4v}$ symmetries. The results are shown in Fig.~\ref{fig:2DHeis_square}(b), where we compare our estimates with projected-pair entangled states (PEPS)~\cite{Liu_2017}, Pixel CNN wave functions~\cite{Sharir_2020}, as well as Quantum Monte Carlo (QMC)~\cite{Liu_2017}. These results show that our ansatz is more accurate compared to PEPS and PixelCNN. We also find that QMC is the best among all the three methods.





\subsection{Heisenberg model on the triangular lattice}

Due to the frustrated, non-bipartite nature of the triangular lattice, the Hamiltonian $\hat{H}$ can no longer be made stoquastic with a simple unitary transformation. Such a Hamiltonian can make the VMC optimization landscape rough and filled with local minima~\cite{Bukov_2021}. Here, we use annealing to overcome local minima and to obtain accurate estimates of ground state energies.

\begin{figure}
    \centering
    \includegraphics[width = 0.95\linewidth]{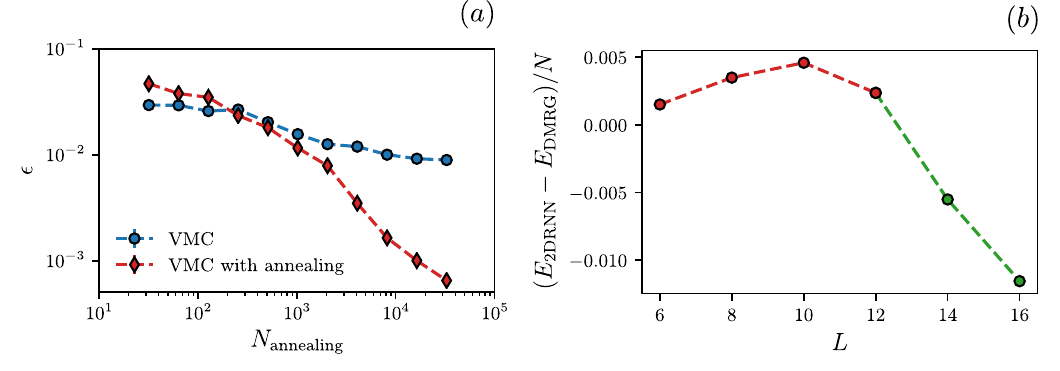}
    \caption{(a) A scaling of the relative error $\epsilon$ against the number of annealing steps $N_{\rm annealing}$ for the triangular Heisenberg model with size $4\times4$, `VMC' corresponds to an initial pseudo-temperature $T_0 = 0$ whereas for `VMC with annealing', we start with $T_0 = 1$. (b) A plot of the energy difference per site between the 2DRNN and the DMRG. Negative values show that our ansatz is superior compared to DMRG for system sizes larger than $14 \times 14$.}
    \label{fig:2DHeis_triangular}
\end{figure}

To demonstrate the idea of annealing, we target the ground state of the $4 \times 4$ triangular lattice at different number of annealing steps $N_{\rm annealing}$. The results in Fig.~\ref{fig:2DHeis_triangular}(a) show that starting from a non-zero $T_0$ allows to obtain a lower value of the relative error as opposed to traditional VMC where the initial pseudo-temperature $T_0 = 0$. We also remark that a higher accuracy is obtained by increasing the number of annealing steps, which underlines the importance of adiabaticity in our scheme.

We now focus our attention on larger system sizes. Here we train our 2D cRNN wave function using annealing for a system size $6\times 6$. We then optimize our ansatz at larger system sizes after increasing the lattice side length by $2$ until we arrive at size $16 \times 16$. This step is done at zero pseudo-temperature and without reintializing the parameters of the RNN. This idea was already proposed in the literature in Ref.~\cite{roth2020iterative}. The latter takes advantage of the translation invariance property encoded in the weight-sharing of the parameters of the RNN (see Eqs.~\ref{eq:2DTRNN},~\ref{eq:prob_softmax} and~\ref{eq:phase_softsign}). The results in Fig.~\ref{fig:2DHeis_triangular}(b) show that for system sizes larger than $14 \times 14$, the variational energy obtained by our RNN ansatz is more accurate compared to the energy obtained by DMRG. Interestingly, our RNN ansatz uses less than $0.1 \%$ of the parameters of DMRG for system sizes larger than $14 \times 14$.


\section*{Conclusion}
In this paper, we have applied a modified variant of the two dimensional RNN wave function~\cite{RNNWF, VNA} for the task of finding the ground state of the 2D Heisenberg model on the square and the triangular lattice. We have shown that applying symmetry and annealing helps to obtain highly accurate ground states. We have also demonstrated the power of our ansatz compared to DMRG for large system size on the triangular lattice, where frustration makes the problem more challenging. We believe that more advanced versions of RNN wave functions along with the idea of applying symmetries and annealing can make our ansatz a highly competive tool for studying quantum many-body systems.

\textbf{Note:} the hyperparameters used to produce the results of this paper are detailed in Appendix.~\ref{sec:hyperparams}.

\begin{ack}
Our RNN implementation is based on Tensorflow~\cite{tensorflow2015-whitepaper} and NumPy~\cite{Harris2020}.
Computer simulations were made possible thanks to the Vector Institute computing cluster and the Compute Canada cluster.  J.C. acknowledges support from Natural Sciences and Engineering Research Council of Canada (NSERC), the Shared Hierarchical Academic Research Computing Network (SHARCNET), Compute Canada, Google Quantum Research Award, and the Canadian Institute for Advanced Research (CIFAR) AI chair program. Research at Perimeter Institute is supported in part by the Government of Canada through the Department of Innovation, Science and Economic Development Canada and by the Province of Ontario through the Ministry of Economic Development, Job Creation and Trade.
\end{ack}

\section*{Broader Impact}
The scope of this paper is within quantum many-body physics, which aims to understand the behaviour of matter based on simple rules at the microscopic scale. In the best case, our method could enhance the state-of-the-art algorithms and enable a better understanding of unknown quantum systems. 

\section*{Checklist}


\begin{enumerate}

\item For all authors...
\begin{enumerate}
  \item Do the main claims made in the abstract and introduction accurately reflect the paper's contributions and scope?
    \answerYes{}
  \item Did you describe the limitations of your work?
    \answerYes{}
  \item Did you discuss any potential negative societal impacts of your work?
    \answerNA{}
  \item Have you read the ethics review guidelines and ensured that your paper conforms to them?
    \answerYes{}
\end{enumerate}

\item If you are including theoretical results...
\begin{enumerate}
  \item Did you state the full set of assumptions of all theoretical results?
    \answerNA{}
	\item Did you include complete proofs of all theoretical results?
    \answerNA{}
\end{enumerate}

\item If you ran experiments...
\begin{enumerate}
  \item Did you include the code, data, and instructions needed to reproduce the main experimental results (either in the supplemental material or as a URL)?
    \answerYes{Appendix.~\ref{sec:GRU} provides more details about the 2DRNN cell we use in our study. We also provide more details in Appendix.~\ref{sec:hyperparams} about the training process to facilitate the reproducibility of our results. Making our code publicly available is still a work in progress.}
  \item Did you specify all the training details (e.g., data splits, hyperparameters, how they were chosen)?
    \answerYes{see Appendix.~\ref{sec:hyperparams}}.
	\item Did you report error bars (e.g., with respect to the random seed after running experiments multiple times)?
    \answerYes{the error bars correspond to a one standard deviation. They are computed after obtaining a large number of samples from the RNN using autoregressive sampling.}
	\item Did you include the total amount of compute and the type of resources used (e.g., type of GPUs, internal cluster, or cloud provider)?
    \answerYes{We provide the type of GPUs we used to produce our results in Appendix.~\ref{sec:hyperparams}. However, the total amount of compute was not included due to the lack of data.}
\end{enumerate}

\item If you are using existing assets (e.g., code, data, models) or curating/releasing new assets...
\begin{enumerate}
  \item If your work uses existing assets, did you cite the creators?
    \answerNA{}
  \item Did you mention the license of the assets?
    \answerNA{}
  \item Did you include any new assets either in the supplemental material or as a URL?
    \answerNA{}
  \item Did you discuss whether and how consent was obtained from people whose data you're using/curating?
    \answerNA{}
  \item Did you discuss whether the data you are using/curating contains personally identifiable information or offensive content?
    \answerNA{}
\end{enumerate}

\item If you used crowdsourcing or conducted research with human subjects...
\begin{enumerate}
  \item Did you include the full text of instructions given to participants and screenshots, if applicable?
    \answerNA{}
  \item Did you describe any potential participant risks, with links to Institutional Review Board (IRB) approvals, if applicable?
    \answerNA{}
  \item Did you include the estimated hourly wage paid to participants and the total amount spent on participant compensation?
    \answerNA{}
\end{enumerate}

\end{enumerate}


\newpage

\appendix

\section{Two dimensional gated recurrent neural networks}
\label{sec:GRU}

In this section, we show how we incorporate the gating mechanism in our 2D cRNN wave function ansatz. The 2D RNN cell we use is based on the following recursion relations:
\begin{align}
    \bm{\tilde{h}}_{i,j} &= \tanh \!\left(
    [\bm{\sigma}_{i-1,j} ; \bm{\sigma}_{i,j-1}]T 
    [\bm{h}_{i-1,j} ; \bm{h}_{i,j-1}] + \bm{b}\right), \\
   \bm{u}_{i,j} &= \text{sigmoid}\!\left(
    [\bm{\sigma}_{i-1,j} ; \bm{\sigma}_{i,j-1}]T_g 
    [\bm{h}_{i-1,j} ; \bm{h}_{i,j-1}] + \bm{b_g} \right), \\
    \bm{h}_{i,j} &= \bm{u}_{i,j} \odot \bm{\tilde{h}}_{i,j} + (1-\bm{u}_{i,j}) \odot (W [\bm{h}_{i-1,j} ; \bm{h}_{i,j-1}]).
\end{align}
Here we obtain the state $\bm{h}_{i,j}$ through a combination of the neighbouring hidden states $\bm{h}_{i-1,j}, \bm{h}_{i,j-1}$ and the candidate hidden state $\bm{\tilde{h}}_{i,j}$. The update gate $\bm{u}_{i,j}$ decides how much the candidate hidden state $\bm{\tilde{h}}_{i,j}$ will be modified. The latter combination allows to circumvent some of the limitations related to the vanishing gradient problem~\cite{zhou2016minimal,shen2019mutual}. Note that the size of the hidden state $\bm{h}_{i,j}$ that we denote as $d_h$ is a hyperparameter that we choose before optimizing the parameters of our ansatz. The tensors $T, T_g$, the weight matrix $W$ and the biases $b, b_g$ are variational parameters of our RNN ansatz in addition to the parameters in Eqs.~\eqref{eq:prob_softmax} and~\eqref{eq:phase_softsign}. 

\begin{figure}
    \centering
    \includegraphics[width = \linewidth]{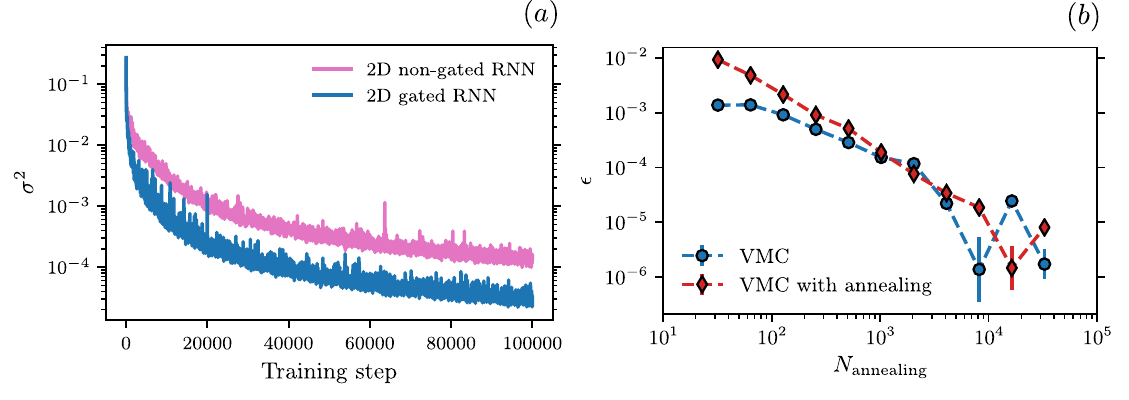}
    \caption{(a) A plot of the energy variance per spin $\sigma^2$ against the number of gradient descent steps, for both a 2D non-gated RNN and a 2D gated RNN. Here we choose the Heisenberg model on a square lattice with size $6\times6$ as a test bed. (b) A scaling of the relative error against the number of annealing steps $N_{\rm annealing}$ for the square Heisenberg model with size $4\times4$, `VMC` corresponds to an initial pseudo-temperature $T_0 = 0$, whereas for `VMC with annealing' we start with $T_0 = 1$.}
    \label{fig:appendix_figure}
\end{figure}

To show the advantage of a 2D gated RNN over the 2D non-gated RNN, we train them using the VMC scheme to find the ground state of the Heisenberg model on the square lattice with size $6\times6$. We find that the gated RNN allows to obtain a more accurate energy. More specifically, we plot the energy variance per spin $\sigma^2 = (\langle \hat{H}^2 \rangle - \langle \hat{H} \rangle)^2/N$ in Fig.~\ref{fig:appendix_figure}(a). This quantity is a good indicator for the proximity of the ansatz to the ground state provided that local minima are avoided~\cite{Claudius90,Assaraf2003,becca2017}. We observe that the gated RNN can get about an order of magnitude lower $\sigma^2$ compared to the non-gated RNN. The latter results demonstrate the advantage of adding the gating mechanism to our wave function ansatz.

\section{Annealing on the square lattice}
\label{sec:annealing}

In the results section, we have used $T_0 = 0$ on the square lattice which is equivalent to not doing annealing. This is motivated by the observation that the Heisenberg model on the square lattice is not frustrated and can made stoquastic using the Marshall sign rule~\cite{Marshall1955, CAPRIOTTI_2001}. To further corroborate this observation, we compare VMC with and without annealing on the square lattice for a system size $4 \times 4$ and report our results in Fig.~\ref{fig:appendix_figure}(b). The latter illustrates that annealing does not show an advantage for large $N_{\rm annealing}$ as opposed to the case of the triangular lattice demonstrated in Fig.~\ref{fig:2DHeis_triangular}(a). This result suggests that in the absence of frustration, there is no need to supplement VMC with annealing.

% ...

\section{Hyperparameters}
\label{sec:hyperparams}

\begin{table*}[p]
    \centering
    \begin{tabular}{|c|c|c|}\hline
       Figures & Parameter & Value \\\hline
        \multirow{5}{*}{Fig.~\ref{fig:2DHeis_square}(a) } & Architecture & 2D Tensorized Gated cRNN wave function \\
            & Number of memory units & $d_h = 300$ \\
            & Number of samples & $N_s = 100$ \\
            & Learning rate & $\eta = 5 \times 10^{-4} \times (1+(t/5000))^{-1}$ \\
            & Number of training steps & $100000$ \\
           \hline
       \multirow{6}{*}{Fig.~\ref{fig:2DHeis_square}(b) } & Architecture & 2D Tensorized Gated cRNN wave function \\
        & Number of memory units & $d_h = 200$ \\
        & Number of samples & $N_s = 100$ \\
        & Learning rate & $\eta = 5 \times 10^{-4} \times (1+(t/5000))^{-1}$ \\
        & Number of training steps & $150000$ \\
        & Applied symmetries & $U(1)$ and $C_{4v}$ \\
       \hline
      \multirow{5}{*}{Fig.~\ref{fig:2DHeis_triangular}(a)} & Architecture & 2D Tensorized Gated cRNN wave function \\
        & Number of memory units & $d_h = 100$ \\
        & Number of samples & $N_s = 100$ \\
        & Learning rate & $\eta = 10^{-4}$ \\
        & Number of warmup steps & $N_{\rm warmup} = 1000$ \\
        & Applied symmetries & $U(1), C_{2d}$ and spin parity \\
        \hline
      \multirow{4}{*}{Fig.~\ref{fig:appendix_figure}(a) }
        & Number of memory units & $d_h = 300$ \\
        & Number of samples & $N_s = 100$ \\
        & Learning rate & $\eta = 5 \times 10^{-4} \times (1+(t/5000))^{-1}$ \\
        & Applied symmetries & $U(1)$ and $C_{4v}$ \\
        \hline 
      \multirow{5}{*}{Fig.~\ref{fig:appendix_figure}(b) } & Architecture & 2D Tensorized Gated cRNN wave function \\
        & Number of memory units & $d_h = 100$ \\
        & Number of samples & $N_s = 100$ \\
        & Learning rate & $\eta = 10^{-4}$ \\
        & Number of warmup steps & $N_{\rm warmup} = 1000$ \\
        & Applied symmetries & $U(1), C_{4v}$ and spin parity \\
        \hline              
    \end{tabular}
    \caption{Hyperparameters used to obtain the results reported in this paper. We note that we use the Adam optimizer to perform gradient descent steps~\cite{kingma2017adam}. }
    \label{tab:hyperparams}
\end{table*}

In this section, we summarize the hyperparameters used to produce the main results of this paper in Tab.~\ref{tab:hyperparams}. We note that the training of our ansatz was performed using P100 GPUs.

In order to produce the results of Fig.~\ref{fig:2DHeis_triangular}(b), we use $d_h = 300$ and $N_s = 100$ samples for training. We have first performed annealing on the system size $6 \times 6$ with $N_{\rm warmup} = 1000$, $N_{\rm annealing} = 10000$, and an initial pseudo-temperature $T_0 = 0.25$, while using a fixed learning rate $\eta = 5\times10^{-4}$. During each annealing step, we perform $N_{\rm train} = 5$ gradient descent steps. In this initial phase, we only apply the U(1) symmetry. In the next phase, we perform an additional $25000$ gradients steps at zero pseudo-temperature and add an additional $25000$ gradients steps after applying $C_{2d}$ symmetry. During this convergence phase, the learning rate is decayed as $\eta = 5\times10^{-5}/(1+t/2000)$, where $t$ corresponds to the current number of convergence steps. We then increase the system size to $8 \times 8$, while keeping our ansatz parameters fixed, applying $C_{2d}$ symmetry and while using zero pseudo-temperature. In this phase, the learning rate is fixed to $\eta = 10^{-5}$. After this step, we train our RNN ansatz until convergence. We repeat the same procedure for system sizes $10 \times 10$, $12 \times 12$, $14 \times 14$ and $16 \times 16$. We note that for $8 \times 8$, we add $40000$ gradient steps. For $10 \times 10$, we continue training with $20000$ gradient steps. For $12 \times 12$, we converge using $10000$ training steps.
For $14 \times 14$, we continue training with $5000$ gradient steps. Finally, for $16 \times 16$, we add $2000$ convergence steps.

To produce the DMRG results in Fig.~\ref{fig:2DHeis_triangular}(b), we used a bond dimension $D = 4000$ for sizes $6 \times 6$, $8 \times 8$ and $10 \times 10$. For $12 \times 12$, we used $D = 3000$ and for the sizes $14 \times 14$ and $16 \times 16$, we used $D = 2000$ since we were not able to obtain the DMRG energy at $D = 4000$ with a limit of $100$ GB memory allocation. The DMRG calculations were run using ITensor~\cite{itensor}.

We finally note that the Marshall sign~\cite{Marshall1955, CAPRIOTTI_2001} is applied on top of our cRNN wave function on the square lattice and on the triangular lattice during all our numerical experiments to speed up convergence. For DMRG, we observed that applying the Marshall sign does not affect the accuracy.

\bibliography{Biblio}

\end{document}